\documentclass[prb]{revtex4}
\usepackage{graphicx}
\usepackage{amsmath}
\usepackage{amsfonts}
\usepackage{amssymb}
\newcommand{\bk}{\boldsymbol{k}}
\newcommand{\bq}{\boldsymbol{q}}
\newcommand{\be}{\begin{equation}}
\newcommand{\ee}{\end{equation}}

\usepackage[usenames]{color}

\begin{document}

\title{Topological Berry phase and semiclassical quantization of
cyclotron orbits\\
for two dimensional electrons in coupled band models}
\author{J.N.~Fuchs, F. Pi\'echon, M.O. Goerbig and G. Montambaux}
\affiliation{Laboratoire de Physique des Solides, Univ. Paris-Sud,
CNRS, UMR 8502, F-91405 Orsay, France.}

\begin{abstract}
The semiclassical quantization of cyclotron orbits for
two-dimensional Bloch electrons in a coupled two band model with a
particle-hole symmetric spectrum is considered. As concrete
examples, we study graphene (both mono and bilayer) and boron
nitride. The main focus is on wave effects -- such as Berry phase
and Maslov index -- occurring at order $\hbar$ in the semiclassical
quantization and producing non-trivial shifts in the resulting
Landau levels. Specifically, we show that the index shift appearing
in the Landau levels is related to a \emph{topological part of} the
Berry phase -- which is basically a winding number of the direction
of the pseudo-spin $1/2$ associated to the coupled bands -- acquired
by an electron during a cyclotron orbit and not to the
\emph{complete} Berry phase, as commonly stated. As a consequence,
the Landau levels of a coupled band insulator are shifted as
compared to a usual band insulator. We also study in detail the
Berry curvature in the whole Brillouin zone on a specific example
(boron nitride) and show that its computation requires care in
defining the ``$\mathbf{k}$-dependent Hamiltonian'' $H(\mathbf{k})$,
where $\mathbf{k}$ is the Bloch wavevector.
\end{abstract}
\date{\today}
\maketitle

\section{Introduction}
The dispersion relation of Bloch electrons in two dimensional (2D)
crystals generally exhibit regions of closed orbits in reciprocal
space. As a consequence, it is expected that applying a
perpendicular magnetic field gives rise to quantized cyclotron
orbits and the corresponding Landau levels. A semiclassical approach
to obtain these Landau levels consists of first computing the area
of the classical cyclotron orbits and then imposing the
Bohr-Sommerfeld quantization condition in the form suggested by
Onsager for Bloch electrons \cite{Onsager}. The semiclassical
quantization condition (see Appendix \ref{app0}) for a cyclotron
orbit $C$ reads:
\begin{equation}
S(C)l_B^2=2\pi[n+\gamma] \label{Onsager}
\end{equation}
where $S(C)\equiv \iint d^2 k$ is the $\boldsymbol{k}$-space area
enclosed by the cyclotron orbit, $\bk$ is the (gauge-invariant)
Bloch wavevector, $l_B\equiv \sqrt{\hbar/eB}$ is the magnetic
length, $-e$ is the electron charge and $n$ is an integer. The
quantity $\gamma$ is called a phase mismatch ($0\leq \gamma<1$) and
is not given by the semiclassical quantization rule \cite{Onsager}.
The precise determination of $\gamma$ requires the inclusion of wave
effects and therefore to include terms of order $\hbar$ in the
semiclassical expansion. For free electrons, and for a single
(uncoupled) band of Bloch electrons, $\gamma=1/2$ as a result of the
presence of two caustics on the cyclotron
orbit\cite{Wilkinson,Keller}. The number of caustics on an orbit is
known as the Maslov index\cite{Keller,Maslov}. From the dependence
of the cyclotron surface $S(C)$ on the energy $\varepsilon$, one can
usually rewrite the above quantization condition as:
\begin{equation}
S(\varepsilon)l_B^2=2\pi[n+\gamma_L]\, . \label{OnsagerEnergy}
\end{equation}
Then by inverting $S(\varepsilon)$, one obtains the (semiclassical)
Landau levels
\begin{equation}
\varepsilon_n=S^{-1}[\frac{2\pi}{l_B^2}(n+\gamma_L)]=\text{function}[B(n+\gamma_L)]
\label{SCLL}
\end{equation}
where $n$ is now interpreted as the Landau index. Usually, the shift
$\gamma_L$ is trivially equal to the phase mismatch $\gamma$
introduced above. For example, the Landau levels for a free electron
of mass $m$ and dispersion relation $\varepsilon=\hbar^2k^2/2m$ are
given by a harmonic oscillator $\varepsilon_n=(n+1/2)\hbar e B/m$,
and $\gamma_L=\gamma=1/2$ in that case. Indeed the area of the
cyclotron orbit is $S(C)=\pi k^2$ and therefore $S(\varepsilon)=2\pi
m \varepsilon/\hbar^2$ such that $S(\varepsilon)l_B^2=2\pi(n+1/2)$.
One of the goal of this paper, it to show that these two quantities,
$\gamma$ and $\gamma_L$, are not necessary equal.

\medskip

A relation between the phase mismatch $\gamma$ and the nature of the
electronic Bloch functions was obtained by Roth \cite{Roth}. She
found that $\gamma$ can depend on the cyclotron orbit $C$ and that
$\gamma(C)$ can be related to a quantity $\Gamma(C)$ later
identified by Wilkinson \cite{Wilkinson} as a Berry phase
\cite{Berry} acquired by the Bloch electron during a cyclotron orbit
$C$, see also Ref. \onlinecite{MS}. The relation reads
\begin{equation}
\gamma(C)=\gamma_M+\gamma_B=\frac{1}{2}-\frac{\Gamma(C)}{2\pi}
\label{RW}
\end{equation}
where $\gamma_M=1/2$ refers to the Maslov index contribution and
$\gamma_B=-\Gamma(C)/2\pi$ to the Berry phase contribution. The
Berry phase is given by
\begin{equation}
\Gamma(C)=i\oint_{C} d\boldsymbol{k} \cdot \langle
u_{\boldsymbol{k}} | \boldsymbol{\nabla}_{\boldsymbol{k}}
 u_{\boldsymbol{k}} \rangle
 \label{BerryPhase}
\end{equation}
in terms of the Bloch function $u_{\boldsymbol{k}}(\boldsymbol{r})$,
where $\boldsymbol{k}$ is the gauge-invariant Bloch wavevector, and
is computed along the cyclotron orbit $C$.

A case which is particularly interesting from the perspective of
semiclassical quantization is that of coupled bands. In the present
paper, we will restrict to two coupled bands with electron-hole
symmetry, having in mind the examples of graphene (the two bands
touch at two inequivalent valleys known as Dirac points) and boron
nitride (the two bands are separated by a gap). In a two coupled
band system, the Bloch electron is endowed with a pseudo-spin $1/2$
associated with the freedom of being in the two bands and its
wavefunction is therefore a bi-spinor. In the context of graphene or
boron nitride, this internal degree of freedom is usual called
``sublattice pseudo-spin'' as it results from having two
inequivalent sites $A$ and $B$ in the unit cell.

The Landau levels of electrons in graphene were first obtained by
McClure \cite{McClure}, who performed a fully quantum mechanical
calculation and obtained the now well-known behavior
\begin{equation}
\varepsilon_{n,\alpha}=\alpha v \sqrt{2neB\hbar} \label{McClureLL}
\end{equation}
where $\alpha=\pm 1$ is the band index and $v$ is the constant Fermi
velocity. From a semiclassical perspective \cite{McClure}, this
result and the value $\gamma_L=0$ that it implies -- via equation
(\ref{SCLL}) -- seem  to imply that the phase mismatch is now
$\gamma=0$ instead of the usual $\gamma=1/2$. Using the
Roth-Wilkinson relation (\ref{RW}), Mikitik and Sharlai \cite{MS}
were able to show that this value of $\gamma$ can be attributed to a
Berry phase $\Gamma=\pi$, which exactly cancels the Maslov
contribution. This Berry phase of $\pi$ is actually the phase
appearing in Hilbert space when rotating a bi-spinor by an angle of
$2\pi$ in real space, which is familiar in the context of spin $1/2$
physics, see e.g. Ref. \onlinecite{Feynman}. The conclusion is that
$\gamma_L=\gamma=0$ in graphene and this value has indeed been
observed in Shubnikov-de Haas and quantum Hall effect measurements
\cite{GeimKimNature2005}.

\medskip

In this paper, we focus on a situation where the assumption that
$\gamma_L=\gamma$ gives a wrong result for the Landau levels. This
is, for example, the case of boron nitride\footnote{More generally,
we consider the case where a gap is opened in graphene's band
structure as a result of inversion symmetry breaking.
Two-dimensional boron nitride is the most obvious example. However
it has a very large gap of roughly $2\Delta \sim 6$~eV, that is even
larger than that of silicon dioxide, and it cannot be used as a
semiconductor\cite{GeimPNAS}. A more interesting system for electronic transport
is graphene on a commensurate (bulk) boron nitride substrate. It is
predicted\cite{Giovannetti} to have a gap $2\Delta \sim 50$~meV. In
the following, we will assume that the gap is smaller
than the bandwidth.}, whose low energy effective theory is that of
massive Dirac fermions \cite{Semenoff}, with a dispersion relation
$\varepsilon(\boldsymbol{k})=\alpha \sqrt{\Delta^2+(v\hbar k)^2}$
where the gap $2\Delta=\varepsilon_B-\varepsilon_N$ is the energy
difference between a boron and a nitrogen $2p_z$ atomic orbital.
Haldane \cite{Haldane} computed the Landau levels of two dimensional
massive Dirac fermions quantum mechanically and found\footnote{This result is valid for
$n>0$. When $n=0$, the result requires the inclusion of both valleys
at the same time: $\varepsilon_{n=0}=-\xi \Delta$, where $\xi=\pm 1$
is the valley index ($K$ or $K'$).}:
\begin{equation}
\varepsilon_{n,\alpha}=\alpha \sqrt{\Delta^2+2neB\hbar v^2}\,\, .
\label{HaldaneLL}
\end{equation}

From a semiclassical perspective Haldane's result raises the
following question. When comparing massless (graphene) and massive
(boron nitride) Dirac fermions, it appears that in both cases, the
Landau level shift is the same $\gamma_L=0$ . However, as we will
show below, the Berry phase $\Gamma(C)$ depends on the magnitude of
the gap and is therefore different in both cases. We are led to
conclude that the two quantities $\gamma$ and $\gamma_L$ are
different in this case. The main goal of this paper is to relate
these two quantities. We will show that while $\gamma$ entering the
quantization (\ref{Onsager}) of cyclotron orbits is correctly
related to the Berry phase, $\gamma_L$ entering the energy
quantization (\ref{OnsagerEnergy}) is related to a topological part
of the Berry phase, which is essentially a winding number of the
pseudo-spin $1/2$. The key point to understand this difference is to
account for the orbital magnetization of Bloch electrons and we will
see that when quantizing the cyclotron orbit one needs to consider
the change in energy due to this magnetization. In the Onsager
quantization condition, the contribution of the orbital
magnetization exactly cancels the non-topological part of the Berry
phase. As a consequence,  the Landau index shift $\gamma_L$ is only
the topological part of the Berry phase and not the whole Berry
phase.


Recently Carmier and Ullmo \cite{CU} have computed the semiclassical
Green functions for similar systems (graphene, boron nitride, etc.).
With a different approach, they reach essentially the same
conclusion as ours: the phase appearing in the Landau levels is in
general not the complete Berry phase but is what they call the
semiclassical phase. In the same vein, see also Ref. \onlinecite{Kormanyos}. 
Another related work, which appeared recently,
is that of Gosselin, B\'erard, Mohrbach and Ghosh \cite{Gosselin2}.
Compared to these two works, on the one hand, ours treats more
general coupled band systems and do not require to take the
continuum limit but also applies to discrete models on a lattice.
For example, the quantization of cyclotron orbits can be performed
anywhere in the Brillouin zone and not only close to specific
points. On the other hand, and contrary to Ref.
\onlinecite{CU,Kormanyos,Gosselin2}, we restrict ourselves to homogeneous
systems.

\medskip

The structure of the paper is as follows. In section II, we review
the semiclassical description of Bloch electrons at order $\hbar$
having in mind the quantization of cyclotron orbits and introduce
various Berry quantities. Section III is the core of the article. It
contains a study of a two coupled band model for which we show that
the Landau index shift $\gamma_L$ is related to a winding number and
not to the complete Berry phase. Then in the following sections, we
consider several examples: a tight-binding model for boron nitride
(IV), massive Dirac electrons (V), massless Dirac electrons (VI),
and eventually chiral electrons of bilayer graphene (VII). The
conclusion is presented in section VIII.

\section{Semiclassical description of a Bloch electron on a cyclotron orbit}
In this section, we review known results about the semiclassical
description (including terms at order $\hbar$) of Bloch electrons in
a crystal under the influence of a magnetic field. When describing a
Bloch electron confined to a single band, the presence of other
bands shows up in the semiclassical equation of motions at order
$\hbar$ (classical order being $\hbar^0$) in the form of Berry phase
type corrections. Our goal is to discuss the effect of these
corrections on the quantization of cyclotron orbits. It should be kept in mind in the following
that the electron is described by a wavefunction which is a bi-spinor -- because of the band structure --
and that the true spin is neglected. For a general review see Ref. \onlinecite{NiuReview}.

\subsection{Semiclassical equations of motion for a Bloch electron in
a magnetic field} One way\cite{Niu,NiuReview} of obtaining the
semiclassical equations of motion for a Bloch electron in a uniform
magnetic field $\boldsymbol{B}$ is to study the motion of a
(typically Gaussian) \emph{wavepacket} of Bloch waves restricted to
a \emph{single band} (indexed by $\alpha$) of average position
$\boldsymbol{r}_c(t)$, average crystal momentum $\hbar
\boldsymbol{q}_c(t)$ -- $\boldsymbol{q}_c$ is the average Bloch
wavevector -- and fixed width. The width of the wavepacket should be
larger than the lattice spacing and much smaller than the typical
length scale on which the external fields (e.g. magnetic and
electric) vary. One then uses the time-dependent variational
principle to obtain an effective Lagrangian for the
\emph{independent} variables $\boldsymbol{r}_c$ and
$\boldsymbol{q}_c$. Minimizing the action with respect to these
variational parameters one obtains the following equations of
motion:
\begin{equation}
\hbar \dot{\boldsymbol{k}}_c=-e\dot{\boldsymbol{r}}_c \times
\boldsymbol{B}\label{scem1}
\end{equation}
and
\begin{equation}
\dot{\boldsymbol{r}}_c=\hbar^{-1}
\boldsymbol{\nabla}_{\boldsymbol{k}_c}\varepsilon_\alpha-\dot{\boldsymbol{k}}_c\times
\boldsymbol{\Omega}_\alpha(\bk_c) \label{scem2}
\end{equation}
where $\hbar \boldsymbol{k}_c\equiv \hbar
\boldsymbol{q}_c+e\boldsymbol{A}(\boldsymbol{r}_c)$ is the average
gauge-invariant crystal momentum\footnote{When studying an electron
in a periodic potential in the presence of a magnetic field, one
should be careful in defining various momenta. Here we consider four
such momenta. First, there is the canonical (or linear) momentum
$\boldsymbol{p}$ which is canonically conjugated to the position
$\boldsymbol{r}$. Second, there is the crystal momentum $\hbar
\boldsymbol{q}$ (where $\boldsymbol{q}$ is the Bloch wavevector)
defined by the Bloch theorem. Third, in the presence of a vector
potential, there is the gauge invariant momentum $\boldsymbol{\Pi}$
which is obtained from the canonical momentum by minimal coupling
$\boldsymbol{\Pi}=\boldsymbol{p}+e\boldsymbol{A}$ (in the absence of
a periodic potential, it is directly related to the velocity
$\boldsymbol{\Pi}=m\boldsymbol{v}$). Fourth, there is the gauge
invariant crystal momentum $\hbar \boldsymbol{k}$, which is related
to the crystal momentum by $\hbar \boldsymbol{k}=\hbar
\boldsymbol{q}+e\boldsymbol{A}$ (in the absence of a magnetic field,
the two are equal $\boldsymbol{k}=\boldsymbol{q}$ and we therefore
usually use $\bk$). The gauge invariant crystal momentum is the one
appearing in the semiclassical equations of motion. It is sometimes
a valid approximation to neglect the difference between linear
momentum and crystal momentum (think of the Peierls substitution),
in such a case $\boldsymbol{p}\simeq \hbar \boldsymbol{q}=\hbar
\boldsymbol{k}-e\boldsymbol{A}$.}, $\boldsymbol{A}$ is the vector
potential and $-e<0$ is the electron charge. The Berry curvature
$\boldsymbol{\Omega}_\alpha(\bk_c)$ is defined below. The electron
energy is
\begin{equation}
\varepsilon_\alpha(\boldsymbol{k}_c)=
\varepsilon_{\alpha,0}(\boldsymbol{k}_c)-\boldsymbol{\mathcal{M}}_\alpha
(\boldsymbol{k}_c)\cdot \boldsymbol{B}
\end{equation}
where $\varepsilon_{\alpha,0}(\boldsymbol{k}_c)$ is the band energy
in absence of a magnetic field and $\boldsymbol{\mathcal{M}}_\alpha
(\boldsymbol{k}_c)$ is the orbital magnetic moment of the Bloch
electron (also defined below).

Compared to the usual equations of motion of Bloch and Peierls
[see e.g. Ref. \onlinecite{AM}] obtained at order $\hbar^0$, there
are two additional terms  in Eq.(\ref{scem1},\ref{scem2}), which
appear at order $\hbar$. One is the so-called anomalous velocity
$-\dot{\boldsymbol{k}}_c\times \boldsymbol{\Omega}_\alpha(\bk_c)$
. It is a kind of Lorentz magnetic force but in
$\boldsymbol{k}$-space and due to Berry curvature
$\boldsymbol{\Omega}_\alpha$, rather than to a real magnetic field.
It takes into account the effect on the average velocity of virtual
transitions to other bands $\alpha'\neq \alpha$. The other is the
magnetization correction to the band energy, which gives the energy
of a Bloch electron in a magnetic field as
$\varepsilon_\alpha=\varepsilon_{\alpha,0}-\boldsymbol{\mathcal{M}}_\alpha
\cdot \boldsymbol{B}$
. The correction to the band energy is the extra magnetic energy due
to the coupling of the orbital magnetic moment
$\boldsymbol{\mathcal{M}}_\alpha(\bk_c)$  to the external magnetic
field. This orbital magnetic moment comes from the
self-rotation\footnote{Note that in addition to the
self-rotation $\boldsymbol{\mathcal{M}}_\alpha(\bk_c)$ the magnetic
moment of the electron also gets a more familiar contribution from
its center of mass motion, which is taken care of by
$\varepsilon_{\alpha,0}(\boldsymbol{k}_c=\bq_c+e\boldsymbol{A}(\boldsymbol{r}_c))\approx
\varepsilon_{\alpha,0}(\bq_c)+e\boldsymbol{A}(\boldsymbol{r}_c)\cdot
\boldsymbol{\nabla}_{\boldsymbol{k}_c}\varepsilon_{\alpha,0}=\varepsilon_{\alpha,0}(\bq_c)+\frac{e}{2\hbar}(\boldsymbol{r}_c\times
\boldsymbol{\nabla}_{\boldsymbol{k}_c}\varepsilon_{\alpha,0})\cdot
\boldsymbol{B}$ in the symmetric gauge where
$\boldsymbol{A}(\boldsymbol{r}_c)=\boldsymbol{B}\times
\boldsymbol{r}_c/2$. The corresponding magnetic moment is
$-\frac{e}{2\hbar}\boldsymbol{r}_c\times
\boldsymbol{\nabla}_{\boldsymbol{k}_c}\varepsilon_{\alpha,0}\approx
-\frac{e}{2}\boldsymbol{r}_c\times \dot{\boldsymbol{r}}_c$ as
expected.} of the wavepacket and exists because of the finite
wavepacket width, which cannot be made arbitrary small due to the
restriction to a single band\cite{NiuReview}. Just as the Berry
curvature, the orbital magnetic moment is also an effect of virtual
transitions to other bands. In the present work, we neglect the
electron spin and therefore do not discuss the Zeeman effect. Note
that the magnetization appears as resulting from an internal
structure, which is not the electron spin but rather the pseudo-spin
related to the two coupled bands.

Berry-type corrections, such as $\Omega_\alpha(\bk_c)$ and
$\boldsymbol{\mathcal{M}}_\alpha(\bk_c)$, appear at first order in
the external field. Therefore, to this order,
$\Omega_\alpha(\bk_c)\approx \Omega_\alpha(\bq_c)$ and
$\boldsymbol{\mathcal{M}}_\alpha(\bk_c)\approx
\boldsymbol{\mathcal{M}}_\alpha(\bq_c)$ and it is therefore not
important to distinguish between $\bk$ and $\bq$ when computing
these quantities.

\subsection{Berry-ology}
Here we consider the effective dynamics of an electron restricted to
a single band and define several quantities related to a Berry phase
appearing because of the coupling between bands.
In particular, we consider a 2D crystalline system described by a
Hamiltonian $\hat{H}$ containing only two bands (band index
$\alpha=\pm 1$). Typically, we think of a tight-binding model with two
sites in the unit cell. In the Bloch basis it reads:
\begin{equation}
\hat{H}=\sum_{\boldsymbol{k},\alpha}\varepsilon_{\alpha,0}(\boldsymbol{k})|\boldsymbol{k},\alpha\rangle
\langle\boldsymbol{k},\alpha|
\end{equation}
where $\varepsilon_{\alpha,0}(\boldsymbol{k})$ is the band energy,
$|\boldsymbol{k},\alpha\rangle = \exp(i\boldsymbol{k}\cdot
\hat{\boldsymbol{r}}) |u_{\boldsymbol{k},\alpha}\rangle $ is a Bloch
state and $\hat{\boldsymbol{r}}$ is the complete position operator
(and not just the Bravais lattice position, e.g.). Its wavefunction
is $\varphi_{\boldsymbol{k},\alpha} (\boldsymbol{r})=\langle
\boldsymbol{r} | \boldsymbol{k},\alpha \rangle =
\exp(i\boldsymbol{k}\cdot \boldsymbol{r})
u_{\boldsymbol{k},\alpha}(\boldsymbol{r})$, where
$u_{\boldsymbol{k},\alpha}(\boldsymbol{r})$ is the Bloch function.
In the case of two bands, $\varphi_{\boldsymbol{k},\alpha}
(\boldsymbol{r})$ and $u_{\boldsymbol{k},\alpha}(\boldsymbol{r})$
are bi-spinors (in sublattice space). Next, we perform a unitary
transform to define a $\boldsymbol{k}$-dependent
Hamiltonian\cite{NiuReview}:
\begin{equation}
\hat{H}(\boldsymbol{k})=\exp(-i\boldsymbol{k}\cdot
\hat{\boldsymbol{r}})\hat{H} \exp(i\boldsymbol{k}\cdot
\hat{\boldsymbol{r}}) \label{ut}
\end{equation}
The wavevector $\boldsymbol{k}$ is a parameter spanning the first
Brillouin zone and on which the Hamiltonian
$\hat{H}(\boldsymbol{k})$ depends. By virtue of the unitary
transform $\exp(-i\boldsymbol{k}\cdot \hat{\boldsymbol{r}})$, one
has $\hat{H}(\boldsymbol{k})|u_{\boldsymbol{k},\alpha}\rangle
=\varepsilon_{\alpha,0}(\boldsymbol{k})|u_{\boldsymbol{k},\alpha}\rangle
$. Using the projection operators
$P({\boldsymbol{k}})=\sum_{\alpha}|u_{\boldsymbol{k},\alpha}\rangle
\langle u_{\boldsymbol{k},\alpha}|$, we also define the following
$2\times 2$ $\boldsymbol{k}$-dependent Hamiltonian:
\begin{equation}
H(\boldsymbol{k})=P(\boldsymbol{k}) \hat{H}(\boldsymbol{k})
P(\boldsymbol{k})=\sum_\alpha
\varepsilon_{0,\alpha}(\boldsymbol{k})|u_{\boldsymbol{k},\alpha}\rangle
\langle u_{\boldsymbol{k},\alpha}|
\end{equation}
which is the restriction of $\hat{H}(\boldsymbol{k})$ to the
$\boldsymbol{k}$ subspace. For more details on the three different types of Hamiltonians we are using [$\hat{H}$, $\hat{H}(\bk)$ and $H(\bk)$] see Appendix \ref{app4}.

Following the general result of Ref. \cite{Berry}, the
Berry phase acquired by a Bloch electron on a cyclotron
orbit\cite{Zak,MS} $C$ in the band $\alpha$ is:
\begin{equation}
\Gamma_\alpha(C)=\oint_{C} d\boldsymbol{k} \cdot i \langle
u_{\boldsymbol{k},\alpha} | \boldsymbol{\nabla}_{\boldsymbol{k}}
 u_{\boldsymbol{k},\alpha} \rangle
\end{equation}
Note that in general this quantity depends on the cyclotron orbit
$C$. The Berry connection (equivalent to a $\boldsymbol{k}$-space
vector potential) in the band $\alpha$ is given by:
\begin{equation}
\boldsymbol{\mathcal{A}}_\alpha (\boldsymbol{k})=i\langle
u_{\boldsymbol{k},\alpha} | \boldsymbol{\nabla}_{\boldsymbol{k}}
 u_{\boldsymbol{k},\alpha} \rangle
 \end{equation}
so that the Berry phase appears as an Aharonov-Bohm phase in
$\boldsymbol{k}$-space. The corresponding Berry curvature
(equivalent to a $\boldsymbol{k}$-space magnetic field) is
\begin{equation}
\boldsymbol{\Omega}_\alpha
(\boldsymbol{k})=\boldsymbol{\nabla}_{\boldsymbol{k}}\times
\boldsymbol{\mathcal{A}}_\alpha = \Omega_\alpha \boldsymbol{e}_z
 \end{equation}
 where
 \begin{equation}
\Omega_\alpha
(\boldsymbol{k})=\partial_{k_x}\mathcal{A}_y-\partial_{k_y}\mathcal{A}_x
=i[\langle
\partial_{k_x} u |  \partial_{k_y} u\rangle - \langle
\partial_{k_y} u |  \partial_{k_x} u\rangle ]
\end{equation}
It can also be written as:
\begin{equation}
\boldsymbol{\Omega}_\alpha (\boldsymbol{k})=i\langle
\boldsymbol{\nabla}_{\boldsymbol{k}}
u_{\boldsymbol{k},\alpha}|\times |
\boldsymbol{\nabla}_{\boldsymbol{k}}
 u_{\boldsymbol{k},\alpha} \rangle
 \label{bc}
 \end{equation}
Another useful formulation, especially convenient when performing
numerical calculations as, contrary to Eq. (\ref{bc}),  it does not
require the Bloch wavefunctions to be single-valued in parameter
space \cite{NiuReview}, is:
\begin{equation}
\boldsymbol{\Omega}_\alpha (\boldsymbol{k})=i\sum_{\alpha' \neq
\alpha}\frac{\langle
u_{\boldsymbol{k},\alpha}|\partial_{k_x}H(\boldsymbol{k})
|u_{\boldsymbol{k},\alpha'} \rangle \langle
u_{\boldsymbol{k},\alpha'}|\partial_{k_y}H(\boldsymbol{k})
|u_{\boldsymbol{k},\alpha}
\rangle}{[\varepsilon_{\alpha,0}(\boldsymbol{k})-\varepsilon_{\alpha',0}(\boldsymbol{k})]^2}
+ \textrm{c.c.}
 \end{equation}
It shows explicitly, that the Berry curvature is due to the
restriction to a single band $\alpha$ and to the resulting virtual
transitions to other bands $\alpha'\neq \alpha$.

The orbital magnetic moment
$\boldsymbol{\mathcal{M}}_\alpha=\mathcal{M}_\alpha
\boldsymbol{e}_z$ of a Bloch electron described by a wavepacket of
average position $\boldsymbol{r}_c$ and average gauge-invariant
crystal momentum $\hbar \bk_c$ restricted to the band $\alpha$
is\cite{Niu,NiuReview}:
  \begin{equation}
\boldsymbol{\mathcal{M}}_\alpha (\bk_c)=-\frac{e}{2m}\langle
(\hat{\boldsymbol{r}}-\boldsymbol{r}_c)\times \hat{\boldsymbol{p}}
\rangle=-i\frac{e}{2\hbar} \langle
\boldsymbol{\nabla}_{\boldsymbol{k}_c} u_{\boldsymbol{k}_c,\alpha} |
\times [\varepsilon_{\alpha,0}-H(\boldsymbol{k}_c)]  |
\boldsymbol{\nabla}_{\boldsymbol{k}_c}
 u_{\boldsymbol{k}_c,\alpha} \rangle
 \label{magnetization}
 \end{equation}
where the average in the first expression is taken over the
wavepacket, $\hat{\boldsymbol{p}}$ is the canonical momentum
operator, and $m$ is the bare electron mass. As the Berry curvature,
this quantity also has an expression revealing the virtual
transitions to other bands:
 \begin{equation}
\boldsymbol{\mathcal{M}}_\alpha
(\boldsymbol{k})=i\frac{e}{2\hbar}\sum_{\alpha' \neq
\alpha}\frac{\langle
u_{\boldsymbol{k},\alpha}|\partial_{k_x}H(\boldsymbol{k})
|u_{\boldsymbol{k},\alpha'} \rangle \langle
u_{\boldsymbol{k},\alpha'}|\partial_{k_y}H(\boldsymbol{k})
|u_{\boldsymbol{k},\alpha}
\rangle}{\varepsilon_{\alpha,0}(\boldsymbol{k})-\varepsilon_{\alpha',0}(\boldsymbol{k})}
+ \textrm{c.c.}
 \end{equation}
This shows that in the case of a single isolated band, both the
Berry curvature and the orbital magnetic moment vanish. Note that
both quantities depend on the off-diagonal (in band index) matrix
elements  $\hbar^{-1}\langle
u_{\boldsymbol{k},\alpha}|\boldsymbol{\nabla}_{\bk}H(\boldsymbol{k})
|u_{\boldsymbol{k},\alpha'} \rangle$ of the velocity operator.

In the particular case of a two-band model with electron-hole
symmetry, the orbital magnetic moment is directly related to the
Berry curvature:
 \begin{equation}
\boldsymbol{\mathcal{M}}_\alpha=
\frac{e}{\hbar}\varepsilon_{\alpha,0} \boldsymbol{\Omega}_\alpha
 \end{equation}
This relation was already obtained in Ref. \onlinecite{Niu2} and we
present a proof in Appendix \ref{app2}.

According to general symmetry arguments \cite{Niu}, the Berry phase
and the magnetization of a single band should vanish in a crystal which
is inversion and time reversal invariant. Indeed, time-reversal symmetry implies
$\boldsymbol{\Omega}(-\boldsymbol{k})=-\boldsymbol{\Omega}(\boldsymbol{k})$
and inversion symmetry implies
$\boldsymbol{\Omega}(-\boldsymbol{k})=\boldsymbol{\Omega}(\boldsymbol{k})$.

All the above definitions are valid for an electron in a single
Bloch band, which is well separated from other bands. We will
nevertheless apply them in the case of touching bands (such as
graphene at its Dirac points) remembering that the correct procedure
is to calculate these quantities in presence of a finite gap
$\Delta$ and to send it to zero at the end.

\subsection{Cyclotron orbit, phase mismatch and Landau index
shift} In the following, the aim is to quantize the cyclotron motion
in order to find the Landau levels. Classically, a free electron in
a uniform and constant magnetic field performs a motion at constant
energy in a plane perpendicular to the magnetic field. For a Bloch
electron, the classical cyclotron orbit is a cut at constant energy
in the band structure, i.e. an iso-energy line
$\varepsilon_\alpha(\boldsymbol{k})=\textrm{constant}$. The
semiclassical quantization of a cyclotron orbit is explained in
detail in the introduction -- see equations (\ref{Onsager}),
(\ref{RW}) and (\ref{BerryPhase}) -- we therefore do not recall it
here. Nevertheless, we would like to precise the definition of the
Landau index shift $\gamma_L$, which is related, but not identical,
to the phase mismatch $\gamma(C)$ appearing in the Onsager
semiclassical quantization condition (\ref{Onsager}). The Landau
index shift appears in the energy quantization condition
(\ref{OnsagerEnergy}). It can also be defined via the exact Landau
levels $\varepsilon_n$ by taking the semiclassical limit ($n\gg 1$,
keeping terms of order $n$ and $n^0$):
\begin{equation}
\varepsilon_n \approx \text{function}[B(n+\gamma_L)]
\end{equation}
where $n$ corresponds to the dominant term, of order $1/\hbar$, and
$\gamma_L$ to the first correction, of order $n^0 \sim 1/\hbar^0$. To be more precise,
imagine expanding the exact Landau levels as a decreasing
series in powers of $n$: $\varepsilon_n=a_0 n^l+a_1 n^{l-1}+a_2
n^{l-2}+\ldots$. Keeping only the two first terms in the
semiclassical limit $n\gg 1$, one obtains $\varepsilon_n\approx a_0
[n^l+a_1 n^{l-1}/a_0]\approx a_0 [n+a_1/(a_0 l)]^l=a_0
[n+\gamma_L]^l$, which defines the Landau index shift
$\gamma_L\equiv a_1/(a_0 l)$ modulo 1.

Often both quantities $\gamma(C)$ and $\gamma_L$ are equal and are
usually not distinguished. The insight here comes from recognizing
that both quantities can be different as $\gamma(C)$ may depend on
the precise cyclotron orbit, whereas $\gamma_L$ is a constant.

\section{Semiclassical quantization of cyclotron orbits in a coupled two-band model}
In the following, we perform the semiclassical quantization of the
cyclotron orbit for a Bloch electron in a two-band model and obtain
the relation between $\gamma(C)$ and $\gamma_L$. We consider a
coupled two-band Hamiltonian with a particle-hole symmetric spectrum
$\hat{H}$. As explained in the previous section, we then perform a
unitary transform $\exp(-i\boldsymbol{k}\cdot \hat{\boldsymbol{r}})$
to obtain a parameter-dependent Hamiltonian
$\hat{H}(\boldsymbol{k})$ and then project on the $\boldsymbol{k}$
subspace to obtain a $2\times 2$ Hamiltonian [in the following
$\hbar\equiv 1$]:
\begin{equation}
H(\boldsymbol{k})=\left(
\begin{array}{cc}\Delta & f(\boldsymbol{k})\\
f^*(\boldsymbol{k}) & -\Delta \end{array} \right)
\end{equation}
where $\boldsymbol{k}$ is the Bloch wavevector in the first
Brillouin zone (BZ). The function $f(\boldsymbol{k})$ is usually
obtained as a sum over hopping amplitudes in a tight binding
description. Time-reversal symmetry imposes
$H(-\boldsymbol{k})^*=H(\boldsymbol{k})$ and therefore
$f(-\boldsymbol{k})^*=f(\boldsymbol{k})$. Note that Bloch's theorem
imposes that
$|f(\boldsymbol{k}+\boldsymbol{G})|=|f(\boldsymbol{k})|$ for any
reciprocal lattice vector $\boldsymbol{G}$. However it does not
require that $f(\boldsymbol{k}+\boldsymbol{G})=f(\boldsymbol{k})$.
An important assumption here is that the diagonal term $\Delta$ does
not depend on the wavevector and can therefore be interpreted simply
as an on-site energy. This term explicitly breaks the inversion
symmetry. Introducing the energy spectrum
$\varepsilon_0(\boldsymbol{k})=\alpha
\sqrt{\Delta^2+|f(\boldsymbol{k})|^2}$, where $\alpha=\pm 1$ is the
band index, and the azimuthal $\beta(\boldsymbol{k})$ and polar
$\theta(\boldsymbol{k})$ angles on the Bloch sphere, such that $\cos
\beta=\Delta/|\varepsilon_0|$, $\sin \beta=|f|/|\varepsilon_0|$ and
$\theta\equiv -\textrm{Arg} f$, the Hamiltonian can be rewritten as
\begin{equation}
H(\boldsymbol{k})=|\varepsilon_0| \left(
\begin{array}{cc}\cos \beta & \sin \beta e^{-i \theta}\\
\sin\beta e^{i\theta} & -\cos \beta \end{array} \right)
\end{equation}
The eigenfunction of energy $\varepsilon_0=\alpha |\varepsilon_0|$
is $\psi(\boldsymbol{r})=u_{\boldsymbol{k}}(\boldsymbol{r})
e^{i\boldsymbol{k}\cdot \boldsymbol{r}}$ where the Bloch spinor is
\begin{eqnarray}
|u_{\boldsymbol{k},\alpha}\rangle &=&
 \left(\begin{array}{c}\cos(\beta/2)\\
 \sin(\beta/2) e^{i\theta} \end{array}\right) \text{ if } \alpha=+1 \nonumber \\
&=&\left(\begin{array}{c} -\sin(\beta/2 )e^{-i\theta}\\
 \cos (\beta/2)\end{array}\right) \text{ if } \alpha=-1
\end{eqnarray}
The Berry connection is given by
\begin{equation}
\boldsymbol{\mathcal{A}}=-\alpha\sin^2\frac{\beta}{2}
\boldsymbol{\nabla}_{\boldsymbol{k}} \theta
\end{equation}
and the corresponding curvature is
\begin{equation}
\boldsymbol{\Omega}=\frac{\alpha}{2}
\boldsymbol{\nabla}_{\boldsymbol{k}} \cos \beta \times
\boldsymbol{\nabla}_{\boldsymbol{k}} \theta=-\frac{\alpha}{2}\sin
\beta (\partial_{k_x}\beta
\partial_{k_y}\theta-\partial_{k_x}\theta
\partial_{k_y}\beta)\boldsymbol{e}_z
\end{equation}

An important simplification occurs in the calculation of the Berry
phase $\Gamma$ because the cyclotron orbit $C$ is travelled at
constant energy and the diagonal term $\Delta$ is independent of the
wavevector. As a consequence, the azimuthal angle $\beta$ is a
constant along the trajectory. Indeed $\cos \beta =
\Delta/|\varepsilon_0|$ and $\sin \beta =
\sqrt{\varepsilon_0^2-\Delta^2}/|\varepsilon_0|$ are both functions
of $\varepsilon_0$ only. Therefore the calculation of the Berry
phase along a cyclotron orbit is easily performed:
\begin{equation}
\Gamma(C)=\oint_C d \boldsymbol{k} \cdot \boldsymbol{\mathcal{A}}=-\alpha
\sin^2\frac{\beta}{2}\oint_C d \boldsymbol{k} \cdot
\boldsymbol{\nabla}_{\boldsymbol{k}} \theta=\pi W_C[1-\cos \beta]
\end{equation}
where $W_C\equiv -\alpha\oint_C d\theta/2\pi$ is the winding number,
which is a topological invariant. Indeed the relevant mapping is from a
cyclotron orbit in the Brillouin zone to a circle (because $\beta$
is fixed) on the Bloch sphere: therefore, the relevant homotopy
group is $\pi_1(S^1)=\mathbb{Z}$. Note that $d (|\varepsilon_0|
\Gamma)/d|\varepsilon_0|=\textrm{constant}=\pi W_C$. We call this
quantity the topological Berry phase. It is a local quantity as it
depends on the precise path $C$. The winding number $W_C$ counts the
total charge of the vortices in $\theta$, which are encircled by the
cyclotron orbit (see Figure 2). Note that this topological Berry phase is not
directly related to the Chern number, which is the Berry curvature
integrated over the entire BZ \cite{TKNN}.


Starting from the Onsager-Roth relation (see Eq.
(\ref{Onsager},\ref{RW},\ref{BerryPhase}))
\begin{equation}
S(\varepsilon_0)l_B^2=2\pi[n+\frac{1}{2}]- \Gamma(C)
\end{equation}
where $\varepsilon_0$ is the band energy in zero magnetic field, we
search the quantization of $S(\varepsilon)$ where $\varepsilon$ is
the energy in presence of a magnetic field. Using the relation
between the energy and the curvature
$\varepsilon_0=\varepsilon+\mathcal{M}B$ with
$\mathcal{M}=e\varepsilon_0 \Omega$, we obtain
\begin{equation}
S(\varepsilon_0)l_B^2=S(\varepsilon)l_B^2+
\bar{\Omega}(\varepsilon_0)|\varepsilon_0| \frac{d
S}{d|\varepsilon_0|}
\end{equation}
In the previous equation, we introduced the Berry curvature
$\bar{\Omega}$ averaged over a constant energy orbit\footnote{When
the dispersion relation $\varepsilon_0(\boldsymbol{k})$ is not
isotropic, the cyclotron orbit in $\boldsymbol{k}$ space is not
circular and the Berry curvature explicitly depends on the
wavevector. Hence the necessity of defining an averaged Berry
curvature. Another expression for this quantity is
$\bar{\Omega}(\varepsilon_0)= [(2\pi)^2\nu(\varepsilon_0)]^{-1}d
\Gamma /d|\varepsilon_0|$ where $\nu(\varepsilon_0)=(2\pi)^{-2}
dS/d|\varepsilon_0|$ is the density of states per unit area.}:
\begin{equation}
\bar{\Omega}(\varepsilon_0)\equiv
\frac{1}{(2\pi)^2\nu(\varepsilon_0)}\frac{d \Gamma}{d|\varepsilon_0|}=\frac{d\Gamma}{dS}
\end{equation}
Therefore, we obtain
\begin{equation}
S(\varepsilon_0)l_B^2=S(\varepsilon)l_B^2+|\varepsilon_0|\frac{d
\Gamma}{d|\varepsilon_0|}
\end{equation}
which does not require the cyclotron orbit to be circular. The
energy quantization condition can now be rewritten as
\begin{equation}
S(\varepsilon)l_B^2=2\pi[n+\frac{1}{2}]-\frac{d (|\varepsilon_0|
\Gamma)}{d|\varepsilon_0|}=2\pi[n+\frac{1}{2}]-\pi W_C \label{main}
\end{equation}
in which we recognized the topological Berry phase. Inverting this last
relation $S(\varepsilon)l_B^2=2\pi[n+(1-W_C)/2]$ allows one to obtain the (semiclassical) Landau levels for
the whole energy band. Finally, the Landau index shift is
\begin{equation}
\gamma_L=\frac{1}{2}-\frac{W_C}{2}
\end{equation}
and the winding number only matters modulo 2. This last equation is
the central result of the paper. It shows that the Landau index
shift $\gamma_L$ is related to the topological part of the Berry
phase $\pi W_C$ and not to the complete Berry phase $\Gamma(C)$. The
important point in the proof is the cancellation in the phase
$S(\varepsilon)l_B^2$ between the non-topological part of the Berry
phase $\Gamma(C)-\pi W_C=-|\varepsilon_0| d\Gamma/d|\varepsilon_0|$
and the orbital magnetic moment contribution
$\mathcal{M}Bd(Sl_B^2)/d|\varepsilon_0|=|\varepsilon_0|
d\Gamma/d|\varepsilon_0|$. Physically, the topological Berry phase
$\pi W_C$ is just the usual $\pi$ phase that a bi-spinor acquires in
Hilbert space as a result of a $2\pi$ rotation in position space.
Here the spin 1/2 is actually the sublattice pseudo-spin.

In the following, we consider several concrete examples such as
boron nitride, graphene mono- and bilayer. These examples are
treated either in discrete lattice models or in their continuum
limit (effective low energy models).

\section{Example 1: tight-binding model of boron nitride}
In this section, we consider a single layer of boron nitride, which
has a honeycomb lattice with two crystallographically and
energetically inequivalent atoms (boron and nitride, usually called
$A$ and $B$) as a basis. Because of the two different on-site
energies $\varepsilon_A-\varepsilon_B=2\Delta\neq 0$, the inversion
symmetry is explicitly broken leading to a gap opening. We use a
tight binding model, with hopping amplitude $t$ and
nearest-neighbour distance $a$, given by the following $2\times 2$
Hamiltonian in $(A,B)$ subspace:
\begin{equation}
H(\boldsymbol{k})=\left(
\begin{array}{cc}\Delta & f(\boldsymbol{k})\\
f^*(\boldsymbol{k}) & -\Delta \end{array} \right) \, \textrm{with}\,
f(\boldsymbol{k})=-t[e^{-i\boldsymbol{k}\cdot
\boldsymbol{\delta}_1}+e^{-i\boldsymbol{k}\cdot
\boldsymbol{\delta}_2}+e^{-i\boldsymbol{k}\cdot
\boldsymbol{\delta}_3}]
\end{equation}
where $\boldsymbol{k}$ is the wavevector in the entire Brillouin
zone [$\boldsymbol{k}=0$ corresponds to the center of the BZ, i.e.
$\Gamma$ point], $\boldsymbol{\delta}_1, \boldsymbol{\delta}_2,
\boldsymbol{\delta}_3$ are vectors connecting an $A$ atom with its
three nearest $B$ neighbours and $\boldsymbol{a}_1,
\boldsymbol{a}_2$ span the Bravais lattice [we follow the notations
of Bena and Montambaux\cite{BenaMontambaux}: our $H(\boldsymbol{k})$
corresponds to what they call basis II\footnote{When computing
the Berry curvature and related quantities, there is no ``choice of
basis'' for $H(\bk)$ in the sense of Ref.
\onlinecite{BenaMontambaux}. Indeed, the basis is fixed by the
unitary transform $\exp(-i\boldsymbol{k}\cdot \hat{\boldsymbol{r}})$
used to define $H(\boldsymbol{k})$ and this gives basis II. Writing
$H(\bk)$ in basis I, instead of II,  amounts to replace $f(\bk)$ by
$f_I(\boldsymbol{k})=-t[1+e^{-i\boldsymbol{k}\cdot
\boldsymbol{a}_1}+e^{-i\boldsymbol{k}\cdot \boldsymbol{a}_2}]$. To
check that this is not correct, we computed the Berry curvature by
boldly replacing $f$ by $f_I$ in the corresponding formulas and
found a different Berry curvature, which did not have the $C_3$
symmetry.}]. Note that, contrary to $|f(\boldsymbol{k})|$,
$f(\boldsymbol{k})$ does not have the periodicity of the reciprocal
lattice but satisfies
$f(\boldsymbol{k}+\boldsymbol{G})=f(\boldsymbol{k})\exp(i\boldsymbol{G}\cdot
\boldsymbol{\delta}_3)$ where $\boldsymbol{\delta}_3$ is the vector
relating the two atoms $A,B$ of the basis. This case exactly
corresponds to that of section III with a specific form for
$f(\boldsymbol{k})$. The quantities of interest (Berry curvature,
orbital magnetic moment, Berry phase, winding number) can be
directly computed from the results obtained there.

\begin{figure}[htb]
\begin{center}
\includegraphics[height=6cm]{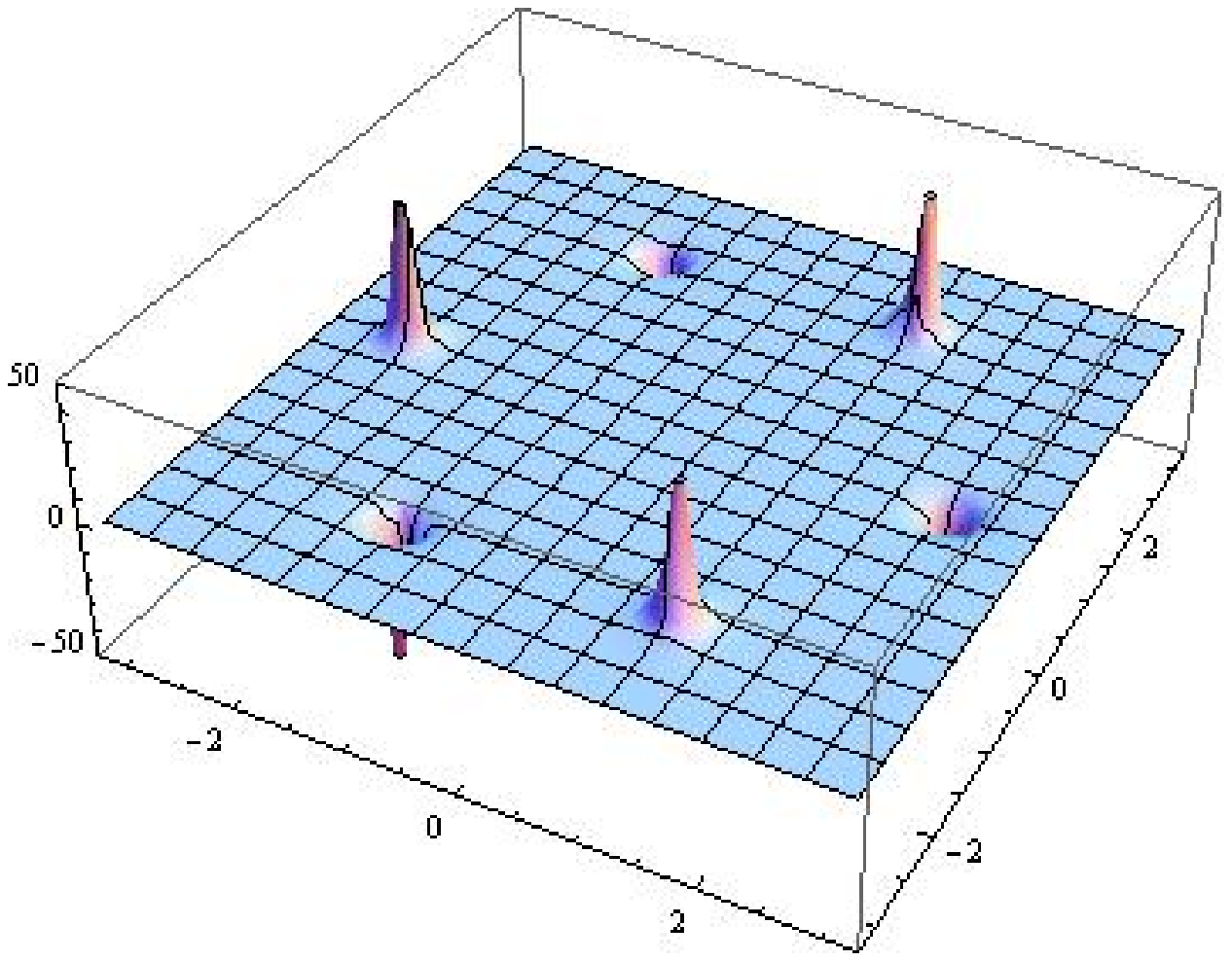}
\includegraphics[height=6cm]{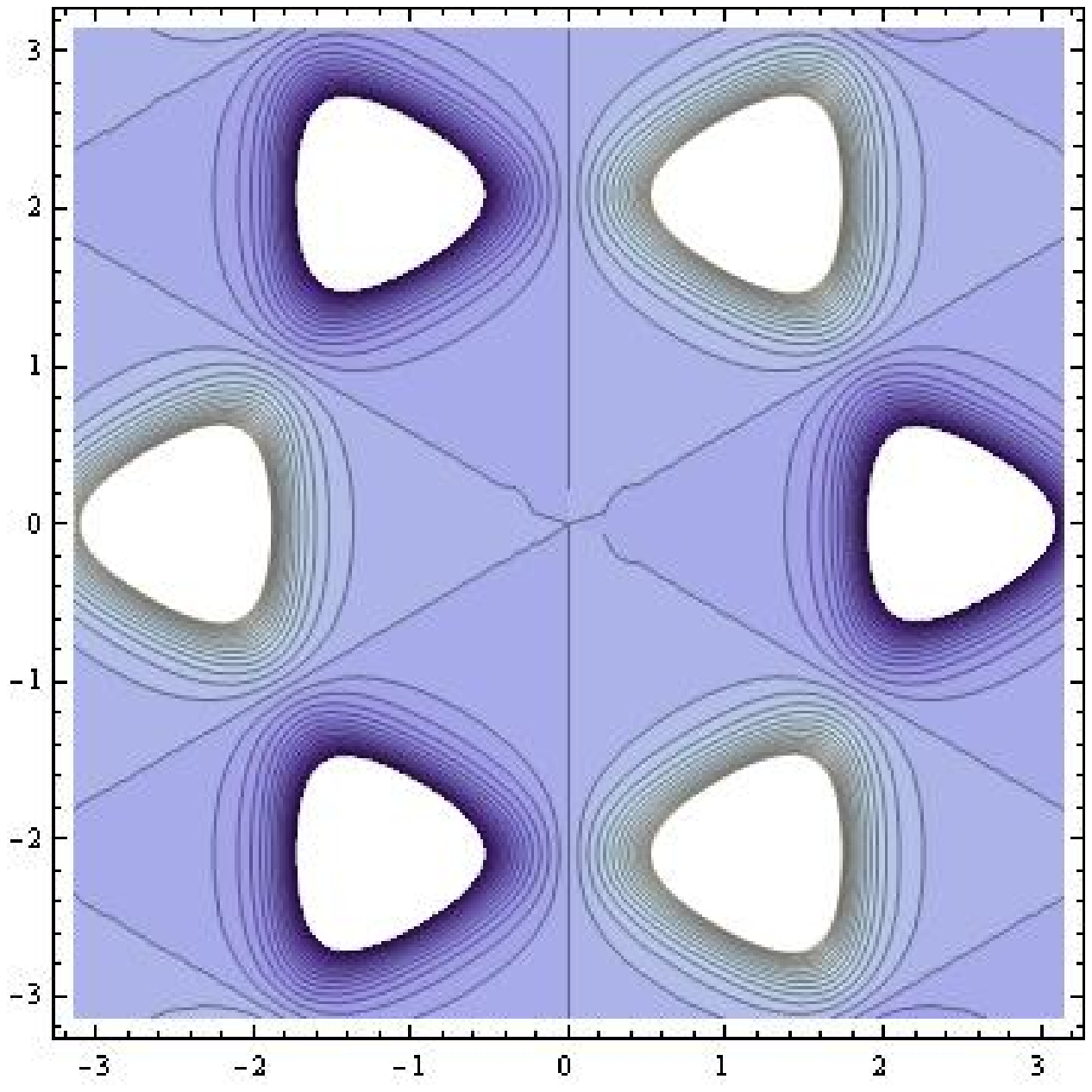}
\label{courburefig} \caption{Berry curvature $\Omega$ [in units of
$a^2$] in the conduction band of boron nitride as a function of the
Bloch wavevector $(k_x,k_y)$ [in units of $1/a$] in the entire
Brillouin zone  for $\Delta/t=0.1$. The lattice vectors have been
taken as
$\boldsymbol{a}_1=\frac{\sqrt{3}}{2}a\boldsymbol{e}_x+\frac{3}{2}a\boldsymbol{e}_y$,
$\boldsymbol{a}_2=-\frac{\sqrt{3}}{2}a\boldsymbol{e}_x+\frac{3}{2}a\boldsymbol{e}_y$.
Left: three dimensional plot $(k_x,k_y,\Omega)$. Right:  contours of
iso-curvature in the Brillouin zone.}
\end{center}
\end{figure}
The curvature is given by
\begin{equation}
\Omega(\boldsymbol{k})
=a^2
\frac{\sqrt{3}\alpha t^2
\Delta}{|\varepsilon_0(\boldsymbol{k})|^3}\sin
(\boldsymbol{k}\cdot\frac{\boldsymbol{\delta}_2-\boldsymbol{\delta}_3}{2})
\sin(\boldsymbol{k}\cdot\frac{\boldsymbol{\delta}_3-\boldsymbol{\delta}_1}{2})
\sin(\boldsymbol{k}\cdot\frac{\boldsymbol{\delta}_1-\boldsymbol{\delta}_2}{2})
\end{equation}
where
$|\varepsilon_0(\boldsymbol{k})|^2=\Delta^2+|f(\boldsymbol{k})|^2$,
see fig. 1. Note that the curvature has both the $C_3$ symmetry and
the translational symmetry
($\Omega(\boldsymbol{k}+\boldsymbol{G})=\Omega(\boldsymbol{k})$)  of
the triangular Bravais lattice.

The orbital magnetic moment is easily obtained from
$\mathcal{M}=e\varepsilon_0 \Omega$ and is shown in fig. 2.
\begin{figure}[htb]
\begin{center}
\includegraphics[height=6cm]{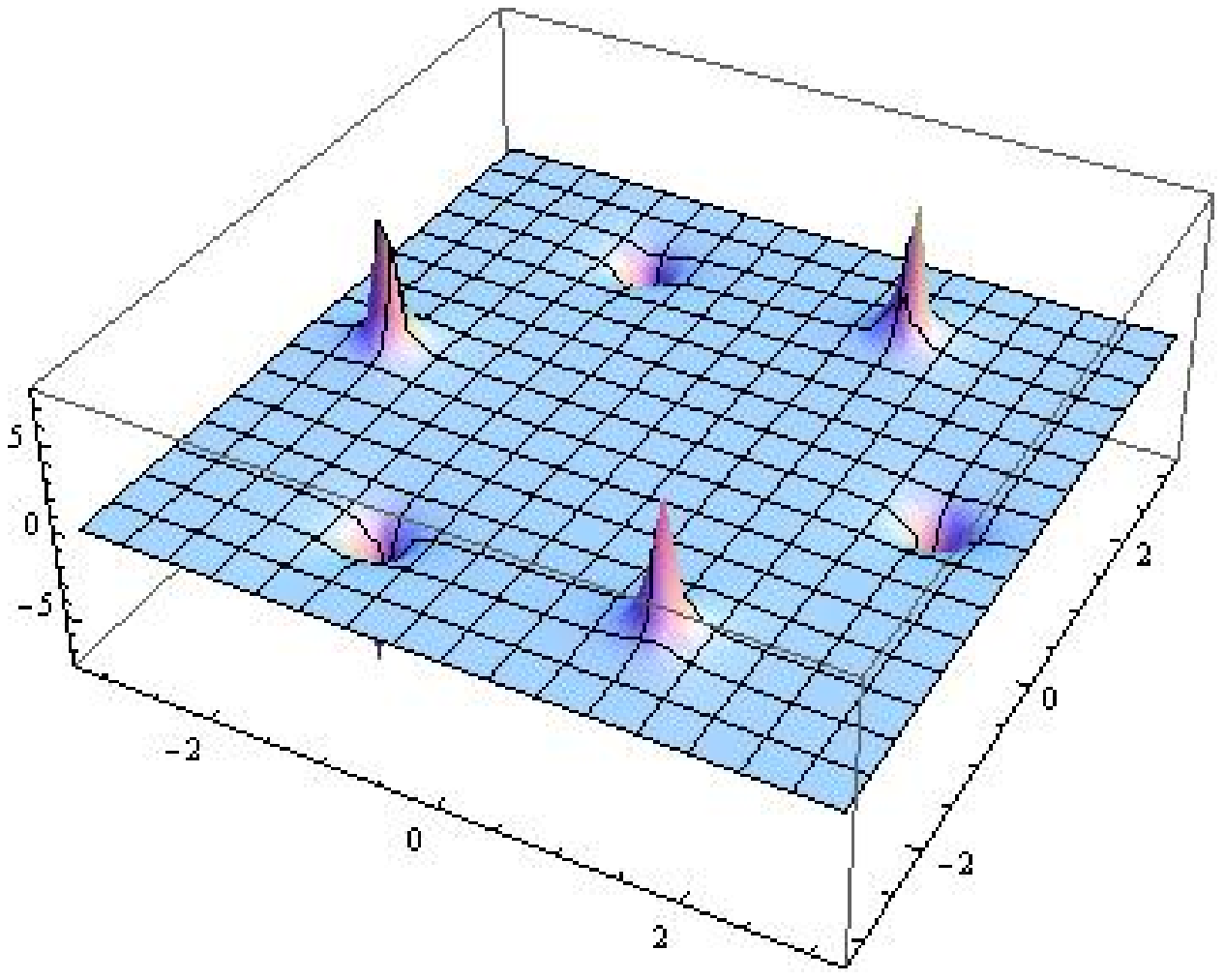}
\includegraphics[height=6cm]{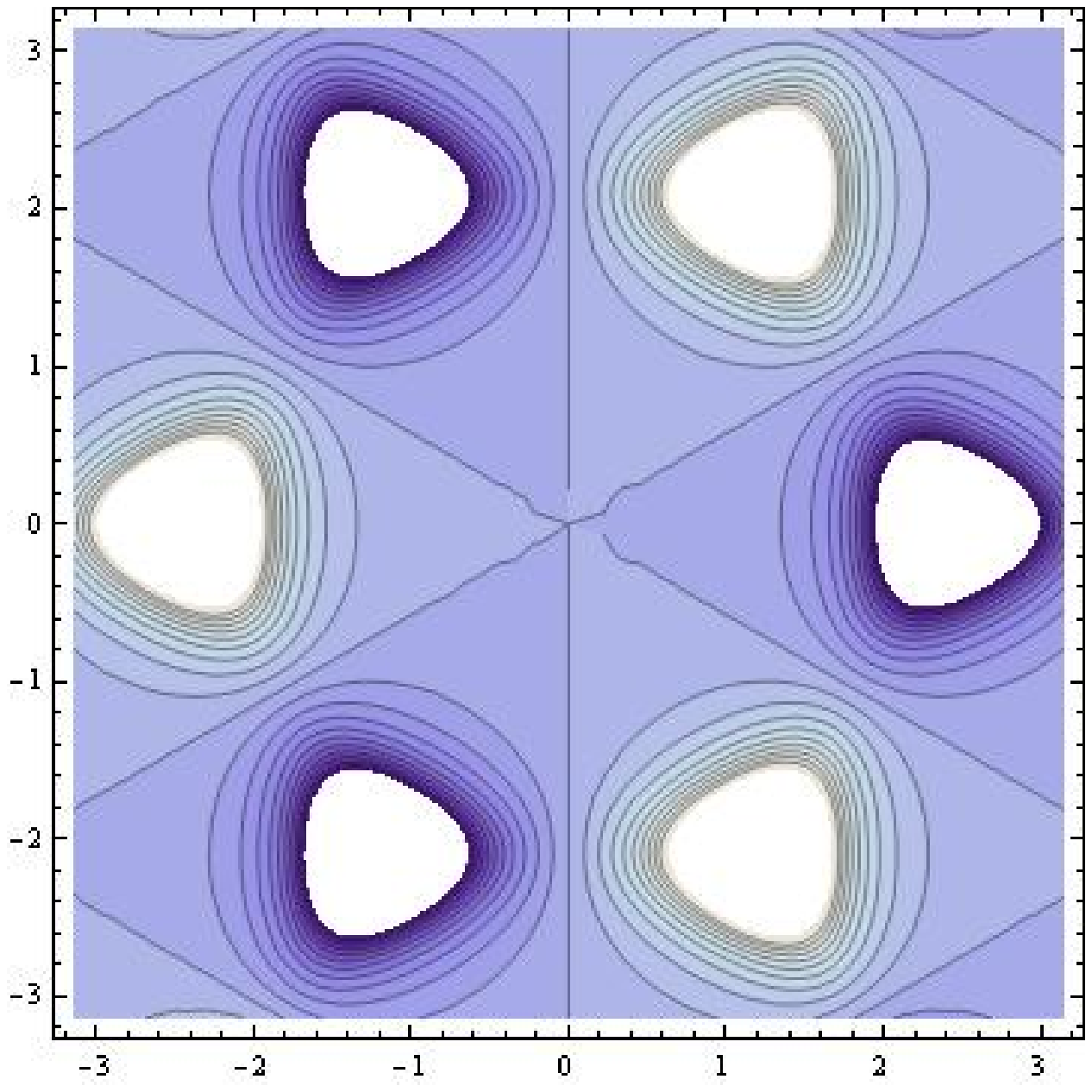}
\label{courburefig} \caption{Orbital magnetic moment
$\mathcal{M}$ [in units of $e\, t\, a^2/\hbar$] in the conduction band of
boron nitride as a function of the Bloch wavevector $(k_x,k_y)$ [in
units of $1/a$] in the entire Brillouin zone  for $\Delta/t=0.1$.
Left: three dimensional plot $(k_x,k_y,\mathcal{M})$. Right:
contours of iso-$\mathcal{M}$ in the Brillouin zone.}
\end{center}
\end{figure}

Because of time reversal symmetry, the curvature satisfies
$\Omega(-\boldsymbol{k})=-\Omega(\boldsymbol{k})$ and its integral
over the entire BZ vanishes. As inversion symmetry is absent
$\Omega(-\boldsymbol{k})\neq \Omega(\boldsymbol{k})$.

The Berry phase for a cyclotron orbit $C$ of constant energy
$\varepsilon_0$ is $\Gamma(C)=\pi
W_C[1-\frac{\Delta}{|\varepsilon_0|}]$ where $W_C\equiv -\alpha
\oint_C d\theta/2\pi$ is the winding number, which is $\pm 1$ when
encircling a valley (because of a vortex in $\theta$) and $0$ when
the orbit is around the $\Gamma$ point, see fig. 3.
\begin{figure}[htb]
\begin{center}
\includegraphics[height=6cm]{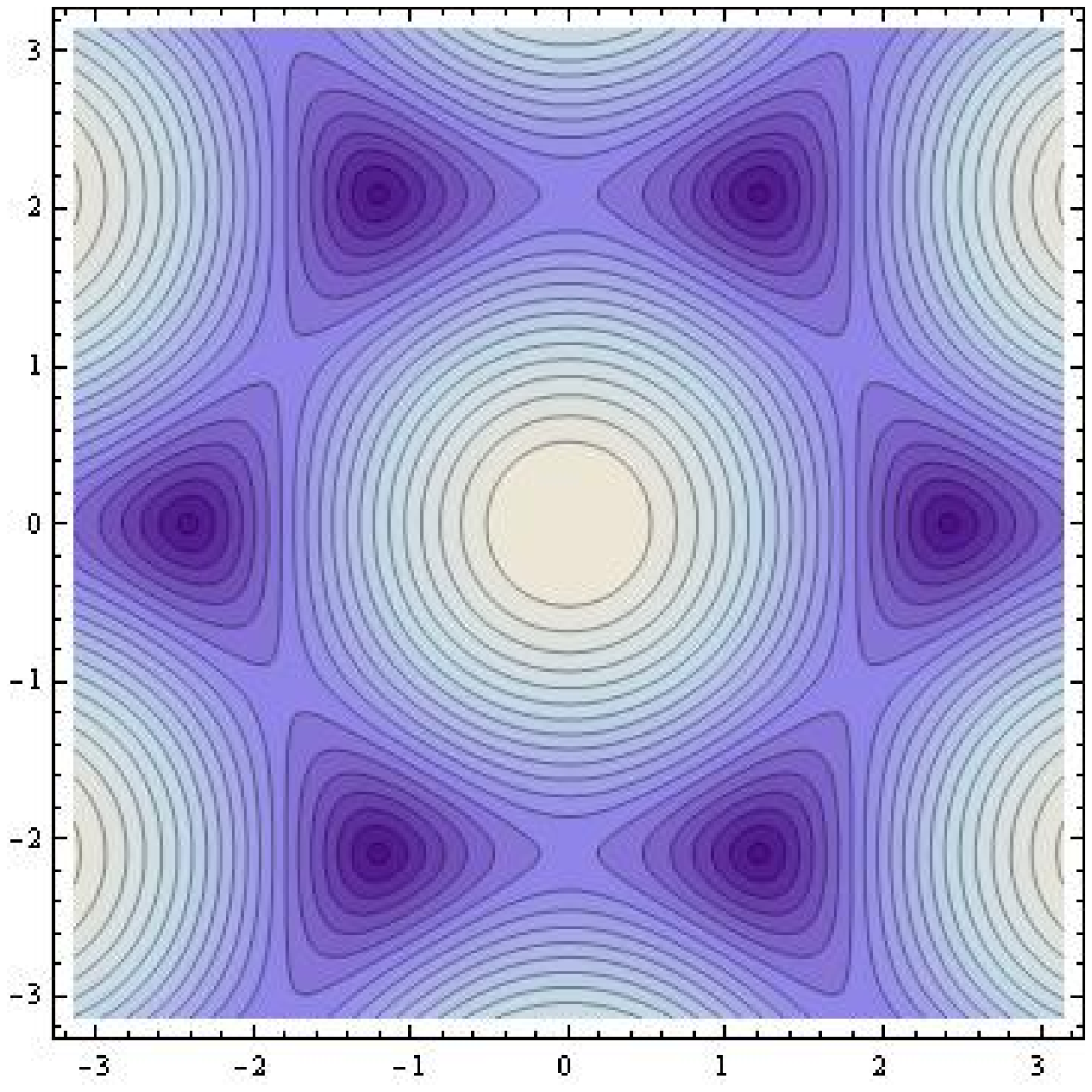}
\includegraphics[height=6cm]{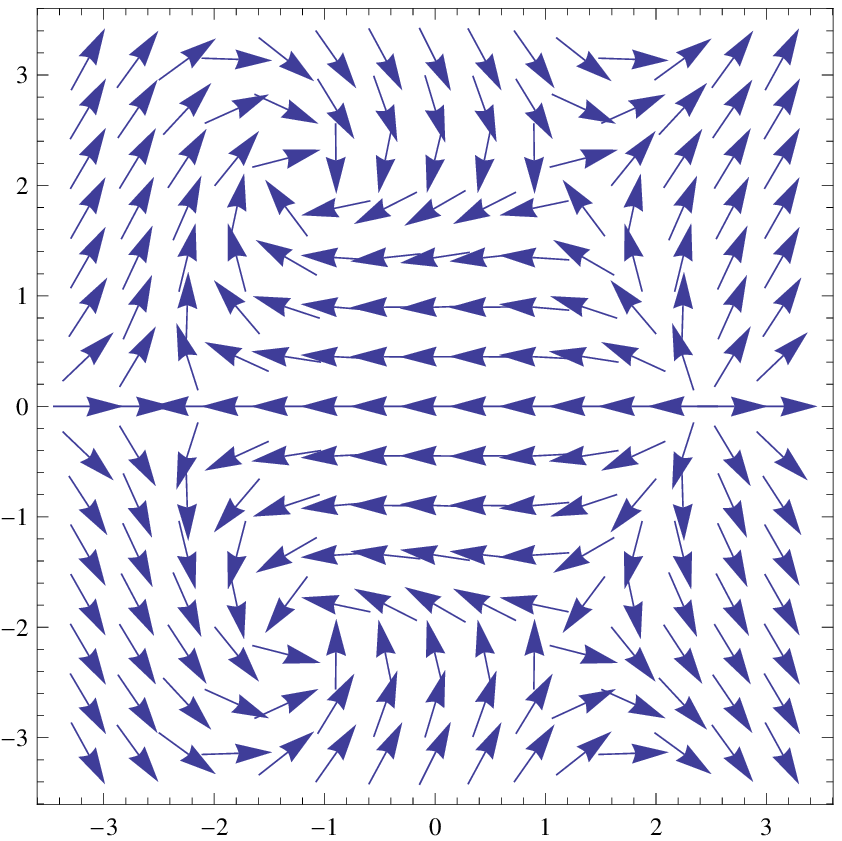}
\label{isoenergies} \caption{Left: Isoenergy lines
($\varepsilon_0(\bk)$=constant) of boron nitride in the first
Brillouin zone for $\Delta=0.1$ [energies in units of $t$]. In the
semiclassical limit, cyclotron orbits in reciprocal space follow the
isoenergy lines. Right: Polar angle on the Bloch sphere
$\theta(\boldsymbol{k})\equiv -\textrm{Arg} f(\boldsymbol{k})$ in
the BZ. The winding number $W_C$ measures the topological charge of
vortices in the polar angle $\theta$.}
\end{center}
\end{figure}
A saddle point in the energy dispersion at
$|\varepsilon_0|=\sqrt{\Delta^2+t^2}$ separates the cyclotron orbits
which encircle the two valleys from the cyclotron orbit which
encircle the $\Gamma$ point in the BZ. As a consequence,
\begin{eqnarray}
\Gamma(C)&=&-\alpha \xi \pi [1-\Delta/|\varepsilon_0|] \,\,\,
\textrm{if} \,\,\, \Delta \leq |\varepsilon_0| <
\sqrt{\Delta^2+t^2} \,\,\, (\textrm{i.e. } W_C=-\alpha\xi=\pm 1)\nonumber \\
&=&0 \,\,\, \textrm{if} \,\,\, \sqrt{\Delta^2+t^2} < |\varepsilon_0|
\leq \sqrt{\Delta^2+(3t)^2} \,\,\, (\textrm{i.e. } W_C=0)
\end{eqnarray}
We checked this simple expression for the Berry phase along a
cyclotron orbit numerically by directly computing the integral of
the curvature in $\boldsymbol{k}$ space over the area encircled by
the cyclotron orbit.

From the energy quantization relation
$S(\varepsilon)l_B^2=2\pi[n+1/2]-\pi W_C$ it is now possible to
obtain the (semiclassical) Landau levels for the whole energy band
of boron nitride. It shows that the Landau index shift $\gamma_L=1/2
\pm 1/2=0$ (modulo 1) vanishes for cyclotron orbits encircling a
single valley ($K$ or $K'$). Whereas for orbits around the $\Gamma$
point, it is $\gamma_L=1/2+0=1/2$.

\section{Example 2: low energy model of boron nitride (massive 2D Dirac fermions)}
We now take the continuum limit of a single layer of boron nitride.
The low energy effective theory close to two inequivalent corners of the Brillouin zone (called
valleys $K$ and $K'$) is now given by a massive 2D Dirac Hamiltonian \cite{Semenoff}:
\begin{equation}
H_\xi (\boldsymbol{k})=\xi \boldsymbol{k}\cdot
\boldsymbol{\sigma}_\xi+\Delta \sigma_z =|\varepsilon_0| \left(
\begin{array}{cc}\cos \beta & \xi \sin \beta e^{-i\xi \theta}\\
\xi \sin\beta e^{i\xi \theta} & -\cos \beta \end{array} \right)
\end{equation}
where $\cos \beta = \Delta/|\varepsilon_0|$, $\sin \beta =
k/|\varepsilon_0|$ with $|\varepsilon_0|=\sqrt{\Delta^2+k^2}\geq 0$
and $0\leq \beta \leq \pi/2$, $\theta(\boldsymbol{k})=\textrm{Arg}
(k_x+ik_y)$ and $f_\xi (\boldsymbol{k})=\xi |\boldsymbol{k}|e^{-i\xi
\theta(\boldsymbol{k})}$. The wavevector $\boldsymbol{k}$ is now defined from the $K$ or
$K'$ points and not in the entire BZ. The Pauli operator vector is defined as
$\boldsymbol{\sigma}_\xi \equiv (\sigma_x,\xi \sigma_y)$ where
$\xi=\pm 1$ is the valley index ($\xi=1$ corresponding to the $K$
valley). The Fermi velocity $v=3ta/2$ has been
taken to 1. The most general
single valued eigenfunction with eigenenergy $\varepsilon_0=\alpha
|\varepsilon_0|$ is
$\psi(\boldsymbol{r})=u_{\boldsymbol{k}}(\boldsymbol{r})
e^{i\boldsymbol{k}\cdot \boldsymbol{r}}$ where the Bloch spinor is
\begin{eqnarray}
|u_{\boldsymbol{k},\alpha}\rangle &=&
 \left(\begin{array}{c}\cos(\beta/2)\\
 \xi \sin(\beta/2) e^{i\xi \theta} \end{array}\right) \text{ if } \alpha=+1 \nonumber \\
&=&\left(\begin{array}{c} -\xi \sin(\beta/2 )e^{-i\xi \theta}\\
 \cos (\beta/2)\end{array}\right) \text{ if } \alpha=-1
\end{eqnarray}

The Berry connection is given by:
\begin{eqnarray}
\boldsymbol{\mathcal{A}}
&=&-\alpha\xi \sin^2(\beta/2)\nabla \theta
\end{eqnarray}
Upon integration over the circular cyclotron orbit $C$ of radius
$k$, we obtain the Berry phase:
\begin{equation}
\Gamma(k)=-\alpha \xi 2\pi \sin^2(\beta/2)=-\alpha \xi \pi(1-\cos
\beta) \label{Gamma}
\end{equation}
and the topological Berry phase:
\begin{equation}
\pi W_C=-\alpha \xi \pi
\end{equation}
The connection can
be rewritten as:
\begin{equation}
\boldsymbol{\mathcal{A}}=
 \frac{\Gamma(k)}{2\pi}\boldsymbol{\nabla}_{\boldsymbol{k}}\theta = \frac{\Gamma(k)}{2\pi k}\boldsymbol{e}_\theta
\end{equation}
Note that
\begin{equation}
\sum_{\xi=\pm} \Gamma_{\alpha,\xi}(k)=0 \label{trc}
\end{equation}
which is a manifestation of time-reversal cancellation. The Berry
phase depends on the magnitude of the gap, which means that it now
depends on $k$ and therefore on the magnetic field $B$.
Two limits of interest are the
``ultra-relativistic'' limit ($\Delta /k\to 0$, $\beta \to \pi/2$)
\begin{equation}
\Gamma(k\gg \Delta)\approx -\alpha \xi \pi=\pi W_C
\end{equation}
and the ``non-relativistic'' limit ($\beta\approx k/\Delta  \to 0$):
\begin{equation}
\Gamma(k\ll \Delta)\approx 0=\Gamma(0)
\end{equation}
when $\Delta\neq 0$. The corresponding Berry curvature is:
\begin{equation}
\Omega=\frac{1}{2\pi k}\frac{d \Gamma}{dk}=-\alpha \xi
\frac{\Delta}{2|\varepsilon_0|^3} \label{curvature}
\end{equation}
It does not contain a singular term, except when $\Delta\to 0^+$:
$\Omega=0$ when $k\neq 0$ and $\Omega \to -\alpha \xi \infty$ when
$k=0$. Details of the calculation are given in Appendix \ref{app3}.

The orbital magnetic moment is \cite{Niu2}
\begin{equation}
\mathcal{M}=e\varepsilon_0 \Omega=-\xi
\frac{e\Delta}{2\varepsilon_0^2} \label{MOmega}
\end{equation}
As a side remark, we note that this orbital magnetic moment leads to
a valley-Zeeman effect in presence of a magnetic field. In
particular, at the bottom of the band $k\to 0$, the orbital magnetic
moment is $\mathcal{M}(0)=-\xi e/2\Delta$ and the valley-Zeeman gap
would be $2\Delta_{vZ}=2\mathcal{M}(0)B$. Some effects related to
this valley magnetic moment are discussed in Ref. \onlinecite{Niu2}.
Here, we would like to point out one more effect, which could be
relevant for graphene in the quantum Hall regime. In graphene --
which is gapless $\Delta=0$ in the absence of a magnetic field -- it
is possible to imagine a self-consistent mechanism at finite $B$
leading to a valley-dependent gap opening for the $n=0$ Landau
level. Indeed asking that the gap leading to a valley magnetic
moment is itself the valley-Zeeman gap $\Delta=\Delta_{vZ}$ leads to
$\Delta=\hbar v/l_B \sqrt{2}\propto \sqrt{B}$. This single electron
mechanism is similar but not identical to that proposed by
Lukyanchuk and Bratkovsky \cite{Lukyanchuk}, as can be seen from the
different magnetic field dependence of the gap (square root versus
linear). A valley splitting of the $n=0$ Landau level of graphene
has indeed been observed in a strong magnetic field \cite{Kim}.
However it is not yet clear what is the relevant microscopic
mechanism (for a review see Ref. \onlinecite{KYang}).

From the Onsager relation and the Berry phase just obtained, we find
the energy quantization condition
$S(\varepsilon)l_B^2=2\pi{n+1/2}-\pi W_C$ with $W_C=-\alpha \xi$.
The area $S(\varepsilon)=\pi[\varepsilon^2-\Delta^2]$ has the same
functional form as $S(C)=\pi[\varepsilon_0^2-\Delta^2]$, but the two
quantities differ by the term $-2\pi \varepsilon_0 \mathcal{M}B$. It
is $S(\varepsilon)$ which is directly related to the Landau levels
[and not $S(C)$]. By inverting $S(\varepsilon)$, the semiclassical
Landau levels are:
\begin{equation}
\varepsilon_n=S^{-1}[2\pi eB(n+\frac{1}{2}-\frac{W_C}{2})]=\alpha
\sqrt{\Delta^2+2eB(n+\frac{1}{2}-\frac{W_C}{2})}
\end{equation}
The energy is therefore quantized as
\begin{equation}
\varepsilon_{n'} = \alpha \sqrt{\Delta^2+eB2n'}
\end{equation}
where $n'=n+(1+\alpha \xi)/2$ is an integer. This result agrees with
the exact expression for the Landau levels (\ref{HaldaneLL}),
including $n'=0$. Indeed, $n'=n=0$ implies $\alpha=-\xi$, which
gives $\varepsilon=\alpha \Delta=-\xi \Delta$.
It is a bit surprising that a semiclassical calculation (including
terms of order $\hbar$) is able to recover exactly a fully quantum
result. This is actually a peculiarity of massive Dirac fermions and
does not occur in more general cases.

\section{Example 3: low energy model of graphene (massless 2D Dirac fermions)}
As another example, we consider the case of graphene, which is a
two-dimensional honeycomb lattice of carbon atoms. It can be seen as the limit
of boron nitride when the gap closes because the two carbon atoms in the unit cell have the same on-site energy.
It is a zero-gap semiconductor and its low energy effective theory --
close to $K$ or $K'$ -- is given by a massless 2D
Dirac Hamiltonian:
\begin{equation}
H_\xi (\boldsymbol{k})=\xi v \boldsymbol{k}\cdot
\boldsymbol{\sigma}_\xi=\xi v k\left(
\begin{array}{cc}0 & e^{-i\xi \theta}\\
e^{i\xi \theta} & 0 \end{array} \right)
\end{equation}
where $\theta =\textrm{Arg} (k_x+i k_y)$ depends on the direction of
the Bloch wavevector $\boldsymbol{k}$ [here defined from the $K$ or
$K'$ points] and
$\xi=\pm 1$ is the valley index ($\xi=1$ corresponding to the $K$
valley). The $2\times 2$ matrix is written in $(A,B)$ space. In the
following, we take the Fermi velocity $v\equiv 1$. The most general
single valued eigenfunction with eigenenergy $\alpha k$ is
$\psi(\boldsymbol{r})=u_{\boldsymbol{k}}(\boldsymbol{r})
e^{i\boldsymbol{k}\cdot \boldsymbol{r}}$ where the Bloch spinor is
\begin{eqnarray}
|u_{\boldsymbol{k},\alpha}\rangle &=& \frac{1}{\sqrt{2}}
 \left(\begin{array}{c}1\\
 \xi e^{i\xi \theta} \end{array}\right) \text{ if } \alpha=+1 \nonumber \\
&=& \frac{1}{\sqrt{2}} \left(\begin{array}{c} -\xi e^{-i\xi \theta}\\
 1\end{array}\right) \text{ if } \alpha=-1
\end{eqnarray}
where $\alpha$ is the band index: $\alpha=+1$ [resp. $-1$]
corresponding to the conduction [resp. valence] band. The first
[resp. second] component of the spinor is the amplitude on the $A$
[resp. B] sublattice for both valleys and the area of the system was
taken as unity.

The Berry connection is given by:
\begin{equation}
\boldsymbol{\mathcal{A}}=
 -\frac{\alpha \xi}{2}\boldsymbol{\nabla}_{\boldsymbol{k}}\theta
 =-\frac{\alpha \xi}{2k}\boldsymbol{e}_\theta
\end{equation}
which shows that it is a pure gauge except for the singularity at
the origin. Because of this vortex, it gives a topological
(quantized) Berry phase:
\begin{equation}
\Gamma=\oint_{C} d\boldsymbol{k} \cdot
\boldsymbol{\mathcal{A}}=-\alpha \xi \pi
\end{equation}
which is independent of the cyclotron orbit. Here the winding number
$W_C=-\alpha\xi=\pm 1$, where $\alpha \xi$ is the chirality of the
massless electron. This allows one to rewrite the Berry connection
as:
\begin{equation}
\boldsymbol{\mathcal{A}}=
 \frac{\Gamma}{2\pi}\boldsymbol{\nabla}_{\boldsymbol{k}}\theta
\end{equation}
The corresponding Berry curvature is singular
\begin{equation}
\Omega=\Gamma \delta^2(\boldsymbol{k}) \label{delta}
\end{equation}
and the Roth-Wilkinson relation (\ref{RW}) between the phase
mismatch $\gamma$ and the Berry phase $\Gamma$ gives:
\begin{equation}
\gamma=\frac{1}{2}-\frac{\Gamma}{2\pi}=\frac{1+\alpha\xi}{2} \equiv
0 \text{ mod. }1
\end{equation}
which is consistent with the Landau levels found by McClure
$\varepsilon_n=\alpha \sqrt{2neB}$. The Berry phase is non-zero here
because of the band degeneracy (Dirac point) and despite the
inversion symmetry being present (which results in $\Omega=0$ in the
absence of band degeneracy\cite{Niu}).
A singular Berry phase in a system with inversion and time-reversal symmetry is a signature of the presence of a Dirac point.

The orbital magnetic moment is also singular:
\begin{equation}
\mathcal{M}=e\Gamma \delta(k) \delta(\theta)=-\alpha \xi \pi e
\delta(k) \delta(\theta)
\end{equation}
However, it plays no role in the quantization of cyclotron orbits
for massless Dirac fermions because the area of the cyclotron orbit
at constant energy $S(\varepsilon)= S(C)-2\pi\varepsilon_0
\mathcal{M}B$ is equal to $S(C)$ as $\varepsilon_0\mathcal{M}\propto
k\delta(k)=0$.

\section{Example 4: low energy model of a gapped graphene bilayer}
The low energy effective theory close to $K$ and $K'$ of a gapped
bilayer graphene is given by the following Hamiltonian
\cite{McCannFalko}:
\begin{equation}
H_\xi (\boldsymbol{k})= \left(
\begin{array}{cc}\Delta & -\frac{\boldsymbol{k}^2}{2m} e^{i\xi 2\phi}\\
-\frac{\boldsymbol{k}^2}{2m} e^{-i\xi 2\phi} & -\Delta \end{array}
\right)
\end{equation}
where $m$ is an effective mass and $\phi = \textrm{Arg }(k_x+ik_y)$.
The function $f_\xi$ is therefore
$f_\xi(\boldsymbol{k})=(-\boldsymbol{k}^2/2m)\exp{(i2\xi\phi)}$,
which shows that $\theta(\boldsymbol{k})=2\xi\phi(\boldsymbol{k}) -
\pi$ and $|\varepsilon_0|=\sqrt{\Delta^2+(\boldsymbol{k}^2/2m)^2}$.

The Berry phase is $\Gamma(C)=\pi W_C[1-\Delta/|\varepsilon_0|]$
where the winding number is:
\begin{equation}
W_C=2\alpha \xi
\end{equation}
because the phase $\theta$ rotates twice as fast as $\phi$. From the
previous analysis of the semiclassical quantization condition
(\ref{main}) including the effect of the orbital magnetic moment, we
find the energy quantization condition
$S(\varepsilon)l_B^2=2\pi{n+1/2}-\pi W_C$. Inverting
$S(\varepsilon)=2\pi m\sqrt{\varepsilon^2-\Delta^2}$, we obtain the
semiclassical Landau levels:
\begin{equation}
\varepsilon_n=\alpha
\sqrt{\Delta^2+\omega_c^2(n+\frac{1}{2}-\alpha\xi)^2}=\alpha
\sqrt{\Delta^2+\omega_c^2(n'+\frac{1}{2})^2}
\end{equation}
where the cyclotron pulsation $\omega_c\equiv eB/m$ and $n'=
n-\alpha \xi$ is an integer. The quantum mechanical
result\cite{McCannFalko} is $\varepsilon_n=\alpha
\sqrt{\Delta^2+\omega_c^2n(n-1)}$, which agrees with the
semiclassical result including the $n^0$ order. Here, however, the
semiclassical results does not match the quantum result to all
orders in $\hbar$.

\section{Conclusion}
We have studied wave effects in the semiclassical quantization of
cyclotron orbits in coupled two-band models, focussing especially on
the case of boron nitride. Two main results of the article are the
following:

First, although the phase mismatch $\gamma(C)$ appearing in the
Onsager quantization condition is related to the complete Berry
phase $\Gamma(C)$, the Landau index shift $\gamma_L$ only gets a
contribution from the topological part of the Berry phase $\pi W_C$
(winding number of the pseudo-spin $1/2$). The latter is a
topological invariant, which allows one to distinguish between two
types of band insulators. On the one hand, zero topological Berry
phase indicates that if inversion symmetry is restored, the bands
are well separated and no Dirac points are present. On the other
hand, a non-zero topological Berry phase is a signature of the
presence of Dirac points in crystals with inversion symmetry.
Therefore a shift in the Landau level index is related to a non-zero
topological Berry phase, which signals the presence of underlying
Dirac points (which are only revealed if inversion symmetry is
restored).

Second, computing the Berry curvature in the entire Brillouin zone
requires care in defining the $\bk$-dependent Hamiltonian. In
particular this Hamiltonian should be written in what Bena and
Montambaux\cite{BenaMontambaux} call basis II and not in basis I,
which is the basis that automatically emerges when performing the
unitary transformation (\ref{ut}). The Berry curvature is a local
physical quantity that in principle could be measured. A challenge
would be to design a ``Berrymeter'' to measure this curvature in the
entire Brillouin zone. An idea would be to measure the anomalous
$g$-factor, which is due to the orbital magnetic moment $\mathcal{M}_\alpha$ and contains
the same information as the Berry curvature. This could be done as a
function of doping -- e.g. electric doping in graphene with a gate
-- giving access to local quantities.

\subsection*{Acknowledgements}
We acknowledge useful discussions with Yshai Avishai, Pierre
Carmier, Pierre Gosselin, Pavel Kalugin, Herv\'e Mohrbach, Denis
Ullmo and Yoshikazu Suzumura.

\appendix
\section{Physical interpretation of the semiclassical quantization condition}
\label{app0} Here we give a physical interpretation of the
semiclassical quantization condition for the cyclotron orbit. These
results are certainly not new, but we collect them because they seem
not to be so well-known. Physically Onsager's quantization is the
condition for the single-valuedness of the semiclassical
wavefunction. It states that the total stationary phase $\phi$
accumulated by an electron around its cyclotron orbit is the sum of
four terms and should equal zero modulo $2\pi$:
\begin{equation}
\phi = \hbar k \times 2\pi r/\hbar -eB\times \pi
r^2/\hbar+\Gamma(k)-\pi=2\pi n \label{phi}
\end{equation}
where $n$ is an integer and $-e<0$ is the electron charge. These
four terms are: the spatial de Broglie phase $k \times 2\pi r$; the
Aharonov-Bohm phase $-eB\times \pi r^2/\hbar$; the Berry phase
$\Gamma(k)$; and the Maslov contribution of $-\pi$.

The two first terms are classical (they arrive at order $\hbar^0$).
The de Broglie phase is just the accumulated phase of a (quasi)
plane wave on a trajectory of length $2\pi r$. For the cyclotron
orbit, because classically $\hbar k = eBr$ as $\hbar
\dot{\bk}=-e\dot{\boldsymbol{r}}\times \boldsymbol{B}$, it can be
rewritten as $eB \times 2\pi r^2/\hbar$. The Aharonov-Bohm phase
comes from the fact that the electron surrounds a region of non-zero
magnetic flux $\Phi=B\times \pi r^2$ and the minus sign comes from
the negative electric charge of the electron. It is given by $-2\pi
\Phi/\Phi_0=-eB\times \pi r^2/\hbar$, where $\Phi_0\equiv h/e$ is
the flux quantum. The Aharanov-Bohm phase can be seen as a Berry
phase due to magnetic curvature in real space. Together these two
terms form the classical reduced action (divided by $\hbar$):
$A_\textrm{cl}/\hbar=\oint \boldsymbol{dr}\cdot
\boldsymbol{p}/\hbar=\pi k^2 \times \hbar/eB=S(k)l_B^2$ where
$\boldsymbol{p}\approx \hbar \bk -e\boldsymbol{A}$.

The two other terms are the first quantum corrections to the
classical action (they appear at order $\hbar$ and represent wave
effects): the Berry phase and the Maslov contribution. The Berry
phase is due to curvature in $\boldsymbol{k}$-space because of the
torus-like topology of the Brillouin zone. It can be seen as an
Aharonov-Bohm phase due to a ``magnetic field in $\boldsymbol{k}$
space'' $\boldsymbol{\Omega}$, whose flux is $\Omega \times \pi
k^2$. The Maslov contribution comes from two caustics (Maslov index
of 2) on the cyclotron orbit, each contributing a factor $-\pi/2$.
The caustics represent singularities in the semiclassical
wavefunction, where the probability density diverges and the phase
picks an extra $-\pi/2$ factor. The caustics are actually not
properties of a single orbit but of a family of classical orbits.
For a detailed discussion of caustics and the extra $\pi/2$ phase
(in the context of optics) see Ref. \onlinecite{Landau2}.

Collecting these four terms, equation (\ref{phi}) can be rewritten as:
\begin{equation}
\frac{A_\textrm{cl}}{\hbar}=S(k)l_B^2=2\pi
[n+\frac{1}{2}-\frac{\Gamma(k)}{2\pi}]
\end{equation}
which is precisely the semiclassical quantization of a cyclotron
orbit including terms of order $\hbar$, with
$\gamma=1/2-\Gamma(k)/2\pi$, see Eq. (\ref{Onsager},\ref{RW}).

\section{Hamiltonians}
\label{app4} In this appendix, we discuss the relation between the
three kind of Hamiltonians used in the main text, namely $\hat{H}$,
$\hat{H}(\boldsymbol{k})$ and $H(\boldsymbol{k})$. Hats are used to
distinguish operators acting in the complete Hilbert space from
those solely acting on band indices.

1) The original Hamiltonian of the system is called $\hat{H}$. In
its eigenbasis of Bloch states it reads: \be
\hat{H}=\sum_{\bk,\alpha}
\varepsilon_{\alpha,0}(\bk)|\bk,\alpha\rangle\langle \bk,\alpha |
\ee

2) Next, we define the unitary operator $\hat{U}(\bk)=\exp(-i\bk
\cdot \hat{\boldsymbol{r}})$ -- where $\hat{\boldsymbol{r}}$ is the
complete position operator and not, for example, merely the Bravais
lattice position operator --, which transforms Bloch states $|\bk,\alpha\rangle$ into
their $u$-part:
$\hat{U}(\bk)|\bk,\alpha\rangle=|u_{\bk,\alpha}\rangle$. This
transformation is just a translation by $\bk$ in reciprocal space.
Performing this unitary transform on $\hat{H}$, we define the
$\bk$-dependent Hamiltonian: \be \hat{H}(\bk)\equiv
\hat{U}(\bk)\hat{H}\hat{U}(\bk)^\dagger=\sum_{\bk',\alpha}
\varepsilon_{\alpha,0}(\bk')\exp(i(\bk'-\bk)\cdot\hat{\boldsymbol{r}})
|u_{\bk',\alpha}\rangle\langle u_{\bk',\alpha} |
\exp(-i(\bk'-\bk)\cdot\hat{\boldsymbol{r}}) \ee It is still an
operator in the complete Hilbert space but it depends on $\bk$ as a
parameter. This transformation actually defines a whole family of
Hamiltonians (one for each wavevector $\bk$ in the Brillouin zone).

3) The $2\times 2$ $\bk$-dependent Hamiltonian is defined as the
restriction of $\hat{H}(\bk)$ on the fixed $\bk$ subspace: \be
H(\bk)\equiv P({\bk}) \hat{H}(\bk) P({\bk})  =\sum_\alpha
\varepsilon_{\alpha,0}(\bk)|u_{\bk,\alpha}\rangle\langle
u_{\bk,\alpha} | \ee where $P({\bk}) \equiv \sum_\alpha
|u_{\bk,\alpha}\rangle\langle u_{\bk,\alpha} |$ are projectors on
the $\bk$ subspace. Note that $H(\bk)$ is only  an operator in band
index subspace. In the case of only two bands, it is therefore a
$2\times 2$ matrix. Note that $H(\bk)$ is not periodic in reciprocal
lattice vectors but its eigenvalues are.

\section{Orbital magnetic moment for an electron-hole symmetric two-band Hamiltonian}
\label{app2} In this Appendix, we prove that there is a simple
relation between the orbital magnetic moment and the Berry curvature
in the case of a two-band model with particle-hole
symmetry\cite{Niu2}. The $2\times 2$ Hamiltonian is
\begin{equation}
H(\boldsymbol{k})= \sum_\alpha
\varepsilon_{0,\alpha}(\boldsymbol{k})P_\alpha(\boldsymbol{k})
 \end{equation}
where $P_\alpha(\boldsymbol{k})\equiv
|u_{\boldsymbol{k},\alpha}\rangle\langle u_{\boldsymbol{k},\alpha}|$
are projectors on each of the two bands (labeled by $\alpha=\pm 1$)
and $\varepsilon_{0,\alpha}(\boldsymbol{k})$ are the band energies.
Particle-hole symmetry together with time-reversal symmetry implies that
\begin{equation}
\varepsilon_{0,-\alpha}(-\bk)=-\varepsilon_{0,\alpha}(\bk)=-\varepsilon_{0,\alpha}(-\bk)
\end{equation}
and the Hamiltonian becomes
\begin{equation}
H(\boldsymbol{k})= \varepsilon_{0,+}(\bk)[P_+(\bk) -
P_-(\bk)]=\varepsilon_{0,-}[P_-(\bk) - P_+(\bk)]
\end{equation}
Using the unit operator in the reduced $\boldsymbol{k}$-space
$I(\boldsymbol{k})=P_+ (\boldsymbol{k}) +P_- (\boldsymbol{k})$, we
can write:
\begin{equation}
\varepsilon_{0,\alpha}(\bk)I(\bk)-H(\boldsymbol{k})= 2\varepsilon_{0,\alpha}(\bk)[I(\bk) -
P_\alpha(\bk)]
\end{equation}
Therefore the orbital magnetic moment (in the upper band, e.g.,
$\alpha=+1$) is
\begin{eqnarray}
\boldsymbol{\mathcal{M}}_+ &=& i\frac{e}{2\hbar} \langle
\boldsymbol{\nabla} u_{+} | \times
[\varepsilon_{0,+}-H(\bk)] | \boldsymbol{\nabla}
 u_{+} \rangle \nonumber \\
 &=& i\frac{e}{\hbar}\varepsilon_{0,+}\langle \boldsymbol{\nabla} u_{+} | \times
[I-P_+] | \boldsymbol{\nabla}
 u_{+} \rangle
\end{eqnarray}
But $\langle \boldsymbol{\nabla} u_{+} | \times I|
\boldsymbol{\nabla}
 u_{+} \rangle = -i \boldsymbol{\Omega}_+$ by definition of the Berry curvature and
 $\langle \boldsymbol{\nabla} u_{+} | \times P_+|
\boldsymbol{\nabla}
 u_{+} \rangle = \langle \boldsymbol{\nabla} u_{+} |u_{+} \rangle \times
 \langle u_{+}|
\boldsymbol{\nabla}
 u_{+} \rangle =\boldsymbol{\mathcal{A}}_+ \times \boldsymbol{\mathcal{A}}_+=0$
 by definition of the Berry connection, therefore
\begin{equation}
\boldsymbol{\mathcal{M}}_\alpha=
\frac{e}{\hbar}\varepsilon_{0,\alpha} \boldsymbol{\Omega}_\alpha
 \end{equation}

\section{Berry curvature of a massive Dirac fermion}
\label{app3} The Berry connection is given by:
\begin{equation}
\boldsymbol{\mathcal{A}}=
 \frac{\Gamma(k)}{2\pi}\boldsymbol{\nabla}_{\boldsymbol{k}}\theta
\end{equation}
Therefore:
\begin{equation}
\boldsymbol{\Omega}=\nabla \times
\boldsymbol{\mathcal{A}}=\frac{\nabla \Gamma(k)}{2\pi} \times \nabla
\theta + \frac{\Gamma(k)}{2\pi} \nabla \times \nabla \theta
\end{equation} Using that $\nabla \times \nabla \theta = 2\pi
\delta^2(\boldsymbol{k})\boldsymbol{e}_z$, we obtain:
\begin{equation}
\Omega=\frac{1}{2\pi k} \frac{d\Gamma(k)}{dk} + \Gamma(0)
\delta^2(\boldsymbol{k})
\end{equation}
Therefore, if $\Delta \neq 0$, $\Gamma(0)=0$ and the Berry curvature
is $\Omega=(d\Gamma/dk)/(2\pi k)=-\alpha \xi \Delta
/(2|\varepsilon_0|^3)$. But if $\Delta=0$, $\Gamma(k)=\Gamma=-\alpha
\xi \pi=\Gamma(0)\neq 0$, $d\Gamma/dk=0$ and the curvature is
$\Omega=\Gamma \delta^2(\boldsymbol{k}) $.

\end{document}